\documentclass[aps,prl,twocolumn]{revtex4}
\usepackage{graphicx}
\begin{document}
\begin{widetext}
\author{E. Orignac}
\title{Disordered Quantum smectics}
\affiliation{Laboratoire de Physique Th{\'e}orique de l'{\'E}cole Normale
  Sup{\'e}rieure CNRS-UMR8549 24,Rue Lhomond 75231 Paris Cedex 05
  France}
\author{R. Chitra}
\affiliation{Laboratoire de Physique Th{\'e}orique des Liquides
CNRS-UMR7600 \\
  Universit{\'e} Pierre et Marie Curie, 4  Place
  Jussieu 75252 Paris Cedex 05 France}

\begin{abstract}
  We study  the impurity pinning  of the Quantum Hall (QH)
  smectic  state arising in two dimensional electron systems in high
Landau levels. We use  replicas and a Gaussian Variational
method to deal with the disorder.  
 The pinned quantum smectic exhibits very anisotropic behaviour, with
  density correlations  along the direction of the stripes 
 manifesting a Bragg-Glass type behaviour i.e.,
 quasi long range order  whereas those in the transverse direction are
infra red divergent. We calculate the 
  dynamical conductivity $\sigma_{xx} ({\bf  q},\omega)$ along the
  stripe direction and find a ${\bf q}$ dependent pinning peak. 
\end{abstract}
\maketitle
\end{widetext}

Two dimensional electron gases subjected to  perpendicular
magnetic fields are known to exhibit a plethora of quantum 
phases. The  Integer and Fractional Quantum Hall
Effects,        
Wigner Crystallization etc\ldots are but some of the phases 
observed experimentally\cite{chakraborty_qhe}.  
Most of these phases appear in strong magnetic fields where  the
electrons occupy the lowest Landau level.
On the other hand, certain anisotropic phases have  been observed in 
 the 2DEG  when the electrons occupy
reasonably high Landau
levels ($\nu=9/2$)
\cite{lilly}.
This seemed to be in accord with  Hartree-Fock
calculations~\cite{cdw},
which predicted the existence of new phases such as bubble phases and charge
density waves (CDW)  for intermediate
to weak magnetic fields. 
Consequently, this experimentally seen anisotropic phase was
thought to be a simple
CDW  where the electron density is modulated in only one
direction   resulting in a ground state where electron rich stripes
alternate with hole rich stripes.  This idea offered a simple explanation of the anisotropic
resistivities measured in experiments.  
Since the CDW  is expected to be 
pinned by the incipient disorder it    implies
the existence of a finite threshold electric field for the CDW to slide.
However, though there was a substantial increase in the differential
conductivity, no well defined threshold was seen 
 in the $I-V$ characteristics of the anisotropic
 phase\cite{lilly}.
This  cast serious doubts on the stability of the CDW phase predicted by
Hartree Fock theory with respect to 
quantum  and thermal fluctuations. It is now believed that  
 rather than a CDW or stripe crystal, the correct stable ground state
 is a fluctuating state analogous to a smectic A liquid
 crystal\cite{fradkin_smectic}. 

In the absence of impurities, the properties of this smectic liquid
crystal have been thoroughly  
investigated 
\cite{macdonald_smectics,barci,coeff,fogler}.
In the presence of impurities, the theoretical situation is much less
clear. Conflicting results on the stability of the smectic phase in
the presence of disorder have been obtained in
\cite{macdonald_smectics} and in \cite{yi_fertig}. In
Ref. \cite{scheidl_smectic},  the effect of disorder on the classical
analog of the quantum Hall smectic  was studied
using the renormalization group 
method developped  in
Ref.~\onlinecite{radzihovsky99_smectique} for the conventional classical smectics.
However, the coupling to disorder used in \cite{yi_fertig,scheidl_smectic} is
chosen in analogy  with the conventional smectic  and is not the pertinent
one for 
the Quantum Hall smectic. 

In the present letter, we revisit the problem of impurity pinning of 
the quantum smectic. We derive the appropriate coupling to impurities
and tackle the effect of disorder non-perturbatively,  using a  
powerful variational method \cite{mezard_variational_replica} that
has previously been applied successfully 
to various quantum elastic systems such as
 dirty Luttinger liquids\cite{giamarchi_columnar_variat} and
Wigner crystals \cite{chitra_wigner}. Our model differs from that
of \cite{yi_fertig,scheidl_smectic}  in the way disorder couples to the system and this difference is crucial in obtaining
the 
transport and correlations in the pinned Quantum Hall smectic.

In the smectic phase, the ground state of the two dimensional electron gas is 
characterized by alternating stripes of filling factors $\nu$ and $\nu+1$ with
a period $\lambda$ which is of the order of a few cyclotron lengths. The
smectic
layers are taken to be parallel to the $x$ direction in the $x-y$ plane.   
These fluctuating stripes interact with each other  on short scales
and are subject to residual  Coulomb long range forces on large scales. 
 Starting from
the theory of \cite{macdonald_smectics} and using a Hamiltonian
framework, we have obtained the following effective action
\cite{chitra_unpublished}  to describe the low energy properties of the
 quantum Hall smectic 
\begin{equation}
  \label{eq:free-action}
  S=\frac 1 \beta \sum_n  \int \frac{d^2\mathbf{q}}{(2\pi)^2}
 G_0^{-1}(\mathbf{q},\omega_n) |\phi(\mathbf{q},\omega_n)|^2,
\end{equation}
\noindent  where $\phi$ is the  coarse-grained local  width of a
stripe and  
\begin{equation}
  \label{eq:def-replica}
 G_0^{-1}(\mathbf{q},\omega_n)=\frac{\omega_n^2}{Qq_x^4+ \frac{\kappa_\perp}{\vert q_x \vert}q_y^2} + \kappa_{\parallel}
    |q_x|.
\end{equation} 
 The coefficients $Q$, $\kappa_\parallel$ and $\kappa_\perp$  can be obtained from the Hartree-Fock
energy calculations on the original electronic problem
\cite{coeff,fertig}. $\omega_{n}= 2\pi n/\beta$ are the standard bosonic
Matsubara frequencies with $\beta = 1/T$. Henceforth, we set $\hbar=1$. 
This action is the same as that derived from  a Chern-Simons
description of the Hall system  presented in
\cite[]{barci}.  In the rest of the Letter, we use the same notations as in
\cite[]{barci}.

In the absence of disorder, the above effective action  leads to the following
correlation functions\cite{barci} for the stripe width, 
$C_\phi(x,0,0) \sim 1/(2\lambda)^2
(\kappa_\perp/\kappa_\parallel)^{1/2} \ln (|x|/\alpha)$ where $\alpha$
is a short distance cutoff and $C_\phi(x,y,\tau)= \langle
(\phi(x,y,\tau)-\phi(0,0,0))^2/2\rangle$. Correlation in the direction
perpendicular to the stripes are infrared divergent.  These correlations
are divergent even  in the case of short ranged forces. Inter-stripes process such as tunneling and
backscattering may render the stripes more rigid in the perpendicular
direction resulting in finite correlations. 
It has however, been shown that coulomb interactions
 do not trigger any instability  
towards Wigner crystallization\cite[]{barci}. This is also 
compatible with the results of Ref.~\onlinecite{fertig} where time
dependent Hartree 
Fock calculations indicate that the smectic state is  indeed stable to quantum fluctuations.  These correlation functions are clearly
anisotropic. Since the density correlation function is related to
the  width correlation by the equation $C_\rho(x,y,0)= \exp{- 2 C_\phi(x,y,0)}$\cite[]{chitra_unpublished,barci}, 
we see that the  density fluctuations of the quantum smectic parallel to the
stripe are  indeed stable to quantum
fluctuations. Within the present formalism, the stripe displacement
and width  being  canonically conjugate variables
~\cite{chitra_unpublished},  correlations of the stripe displacement 
can also be calculated.

Interaction of the system with impurities correspond to backward scattering of
chiral electrons or holes  from one  edge of the QH stripe to  another edge of opposite
chirality. Disorder induces both intra-stripe 
and  inter-stripe backward scattering processes. We shall assume in the
following that the intra-stripe processes are the dominant ones and neglect
inter-stripe processes. This
approximation is valid for inter-stripe distances much larger than the
stripe width. In the continuum limit, the  coupling of disorder to the
electron density 
can be written using the expression for 
the edge fermion creation and annihilation
operators\cite[]{macdonald_smectics}. In terms of the field $\phi$, 
the combined action  reads:
\begin{equation}
\label{eq:disordered-action}
  S=S_0 +  \int d^2\mathbf{r}d\tau \left[ \xi(\mathbf{r}) e^{2i
      \phi(\mathbf{r},\tau)} + \text{H. c.}\right]
\end{equation}
\noindent  
For tractability, we assume  a Gaussian distributed disorder,
$\overline{\xi(\mathbf{r}) \xi^*(\mathbf{r'})}=\frac \Delta 4
\delta(\mathbf{r}-\mathbf{r'})$ where $\Delta$ is a measure of the
strength
of  disorder. 
In Ref.~\cite{macdonald_smectics}, it has been shown that this type of
disorder is a relevant perturbation. 
In Refs.~\cite{yi_fertig,scheidl_smectic}, 
 it was incorrectly 
assumed that
disorder coupled to the average displacement of the stripe
 instead of the electron density. In physical terms, the effect of impurities is not to deform the
stripe structure by coupling to the position of the stripe as in the
classical case, but
rather to cut the stripes into  disconnected pieces thus suppressing
transport along the stripe direction. This difference of coupling is
the reason for the discrepancies of \cite{macdonald_smectics} and
\cite{yi_fertig}.

Before, we proceed with the full treatment of the disordered Hall smectic, 
we first look at the equivalent
of the Larkin random force model~\cite{larkin_70} for a $d$-dimensional quantum  smectic (consisting of $d-1$ dimensional
stripes stacked along the $y$ axis) to estimate the relevance of disorder.  
  Within the Larkin
approximation, where the disorder couples directly to the width $\phi$, the width correlation function is given
by 
\begin{equation}
C_\phi (x,y,0)= \frac{\Delta}{(2\pi)^d}\int_0^\infty d^{d-1}q_x \int_{-\frac{\pi}{\lambda}}
^{\frac{\pi}{\lambda}} dq_y    \frac{(1-\cos({\bf q}_x\cdot {\bf  x}+q_yy))}{   \kappa_{\parallel}^2 q_x^2},
\end{equation}
\noindent 
which yields  the following correlations:
$C_\phi(x,0,0) \sim \frac{\Delta }{\kappa_\parallel} x^{3-d}$, with
fluctuations in the perpendicular direction remaining divergent. 
From this analysis, smectic order along
the stripes is very sensitive to dimensionality. 
The upper critical dimension above which disorder is irrelevant  is
$d=3$ and the lower critical
dimension is $d=2$ below which, quantum and thermal  fluctuations effectively
wipe out
the disorder. 
Thus, the quantum smectic is affected by impurities in a strongly
anisotropic manner. Note  that here we are
dealing exclusively with the effect of disorder on  the correlations
of the stripe width and the conclusions do not hold for the stripe
displacement
correlations. However, a similar analysis can be done to study the
effect of disorder on the correlation of the stripe  displacements
\cite{chitra_unpublished}.

The Larkin analysis can also be used to estimate the length scale
$R_a$ over which 
the system  disorders. In the present problem, since the field $\phi$ is like
a phase variable, $R_a$ is defined as the length scale over which the
relative phase changes by $2\pi$ i.e.,
$C_\phi(R_a,0,0)= (2\pi)^2$.
We find that in $d=2$, 
\begin{equation}\label{eq:Ra}
R_a \simeq \frac{ 4 \pi^2 \kappa_{\parallel}^2 \lambda }{\Delta}
\end{equation}
  To go beyond the Larkin approximation,   we now use replicas and the Gaussian
variational method to study the long wavelength properties of the Quantum Hall smectic with both
Coulomb repulsion and impurities described by the action
(\ref{eq:disordered-action}). 
Introducing replicas and averaging over disorder, we obtain the following effective action, 
\begin{widetext}
\begin{equation}
  \label{eq:sreplicated}
  S_{\text{r}}=\sum_a S_0[\phi_a]-\frac{\Delta}{4} \sum_{ab} \int dx dy d\tau
  d\tau' \cos 2[\phi_a(x,y,\tau)-\phi_b(x,y,\tau')] 
\end{equation}
\end{widetext}
To deal with this effective action where the disorder manifests itself as
an interaction non-local in time, we use the  Feynman variational principle.
We introduce 
the  following quadratic variational action:
 \begin{equation}
   \label{eq:svaria}
   S_{\text{trial}}= \frac 1 {2\beta}  \sum_n \int
   \frac{d^2\mathbf{q}}{(2\pi)^2} \sum_{a,b} G_{ab}(\mathbf{q},i\omega_n)
   |\phi(\mathbf{q},i\omega_n)|^2 
 \end{equation}
\noindent where the trial Green's functions $G_{ab}$ are parameterized as $  G_{ab}^{-1}(\mathbf{q},i\omega_n)=G_0^{-1}(\mathbf{q},i\omega_n)\delta_{ab} -\sigma_{ab}(\mathbf{q},i\omega_n) $
  and
$\sigma_{ab}$ are the disorder induced replica self energies.  
Since the disorder is local, the
$\sigma_{ab}$ are independent of the momentum $\mathbf{q}$. These
self-energies are determined in a self-consistent manner by extremizing
the variational free energy
$F_{\text{var}}=F_{\text{trial}} +\frac 1 \beta \langle S_{\text r}-S_{\text{trial}}
\rangle_{S_{\text{trial}}}$ with respect to the $G_{ab}$.
 There are two generic solutions to these
  saddle point equations, one with replica symmetry where
  $\sigma_{a\ne b}=\sigma, \forall a,b$ and  the other with broken
  replica symmetry.    For the quantum system studied in this letter the
  appropriate low temperature solution is the one with replica
  symmetry breaking \cite{giamarchi_columnar_variat}. 

Using the machinery developed in
\cite{mezard_variational_replica} and proceeding 
along the lines of \cite{giamarchi_columnar_variat}, we find that a full
treatment of the disorder in a  $d$ dimensional smectic,
 confirms the Larkin analysis predictions that
$d=3$ ($d=2$) are the upper (lower) critical dimensions for disorder 
~\cite{chitra_unpublished}.
Moreover, since $d=2$ is the lower critical dimension,
the solution to the problem of the Hall smectic considered here, has
one-step replica symmetry breaking. For the one-step solution, 
 the  connected part of the Green's function   defined as
$G_c^{-1}= \sum_b G_{ab}^{-1}$  has the following structure:
\begin{equation}
  \label{eq:varia-1step}
  G_c^{-1}(q,i\omega_n)=G_0^{-1}(q,i\omega_n)+I(i\omega_n)
+\Sigma(1-\delta_{\omega_n,0}). 
\end{equation}
\noindent $\Sigma$  which is independent of both $q$ and the frequency,
 is a mass  term describing impurity pinning
while $I(i\omega_n)$ describes the  dissipation induced by disorder.
The saddle-point equations can be reduced~\cite{chitra_unpublished} to
self-consistent equations $\Sigma$ and $I(i\omega_n)$.
 We find that $\Sigma = \kappa_{\parallel} R_a^{-1}$ with $R_a$ given
by (\ref{eq:Ra}) \cite{chitra_unpublished}.
 $I$ is self consistently determined in the semi-classical
limit \cite{giamarchi_columnar_variat} by the equation
\begin{widetext}
\begin{equation}\label{eq:I-omega}
I(i\omega_n) =\frac { \lambda \kappa_\parallel \Sigma} {4\pi}\int_{-\infty}^{\infty} dq_x
\int_{-\frac{\pi}{\lambda} }^{\frac{\pi}{\lambda} } dq_y \left[\frac
  {1}{\kappa_{\parallel}\vert q_x \vert +\Sigma} -\frac{ Q q_x^4 +
    \frac{\kappa_\perp}{\vert q_x \vert}q_y^2}{\omega_n^2
    +(\kappa_{\parallel} \vert q_x\vert + \Sigma +I (i\omega_n))(Q q_x^4 + \frac{\kappa_\perp}{\vert q_x \vert}q_y^2)}\right]
\end{equation}
\end{widetext}
\noindent 
$I$ determines all dynamical properties and for $\omega_n \to 0$: 
$I(i\omega_n) \sim \omega_n^\frac78$.
 For arbitrary  frequencies, the equation (\ref{eq:I-omega})
can be
continued to real frequencies and  solved  numerically.  

We now study the effect of disorder on various physical properties at
zero temperature. 
Using the results of \cite[]{giamarchi_columnar_variat}, the disorder
averaged equal-time width correlation function 
 along the stripe, is obtained as:
\begin{equation}\label{eq:width-flukes} 
{\tilde C}_\phi(x,0,0) \simeq \frac1{2\pi\lambda \kappa_{\parallel}} 
\log\left(\frac{ x}
{R_a}\right)
\end{equation}
The system retains logarithmic correlations along the stripe
directions, albeit with a different prefactor, which is now
independent of $\kappa_\perp$. Also, the short distance cutoff
$\alpha$ is replaced by the lengthscale $R_a$. Correlations in the
perpendicular direction remain  infra-red divergent as before. Logarithmic
increase of correlations in disordered system are characteristic of
Bragg-Glass phases where density correlations have quasi-long
range order.
Similar Bragg-glass phases have  been previously obtained in  Wigner
crystals\cite{chitra_wigner} 
and vortex lattices\cite{giamarchi_columnar_variat}.
Although these density fluctuations are what is measured
by  diffraction, 
the true order of the stripe phase is
measured by correlation of the displacement. Since the displacement
and width fluctuations are  
canonically conjugate, pinning of the width fluctuations   would imply
that the stripe structure becomes strongly disordered by wild fluctuations
in the displacement.
Note that our results differ greatly from that of \cite{scheidl_smectic} where
logarithmic correlations were obtained for the stripe displacements in both
parallel and perpendicular directions. This can be attributed to the fact
that in our case disorder couples to the density whereas in
Ref.~\cite{scheidl_smectic}, disorder couples to the stripe
displacement and the all important quantum fluctuations have been neglected.
 
Although disorder does not modify strongly the static correlations of
the system, it modifies the transport properties in a nontrivial
manner. To discuss the transport properties, we  need
to calculate the conductivity tensor 
$\sigma_{ij}({\bf q},\omega) = \frac{i}{\omega} \langle J_i({\bf q},
\omega) J_j(-{\bf q},
-\omega) \rangle$.
\noindent
where $\omega$ is the  frequency.
Clearly, in such an anisotropic state the
conductivities parallel and perpendicular to the stripes are
unequal. In addition, due to the presence of an external magnetic
field, there is the Hall conductivity. Within our approach, we can  only access
the diagonal component $\sigma_{xx}$ of the conductivity tensor. 
To obtain the current $J_x$ along the
$x$ direction, we use the equation of continuity which for for $q_y=0$
leads to the expression $q_x J_x(q_x, q_y=0,\omega) = \omega
\rho(q_x,q_y=0,\omega)$ where the density  $\rho(q_x,q_y=0,\omega)= i q_x
\phi(q_x,q_y=0,\omega)$.  
Using this expression for the current,  and analytically continuing the
Matsubara frequencies to real frequencies ($i\omega_n \to \omega + i \epsilon$)in  (\ref{eq:varia-1step}), the conductivity along the
$x$ direction is given by 
\begin{equation}\label{conductivity}
\sigma_{xx}(q_x,q_y=0,\omega)= Ci\omega \frac{Qq_x^4}{\omega^2 -
(\kappa_{\parallel}\vert q_x\vert + \Sigma +I) Q q_x^4}
\end{equation}

\begin{figure}
\centerline{\includegraphics[width=8cm]{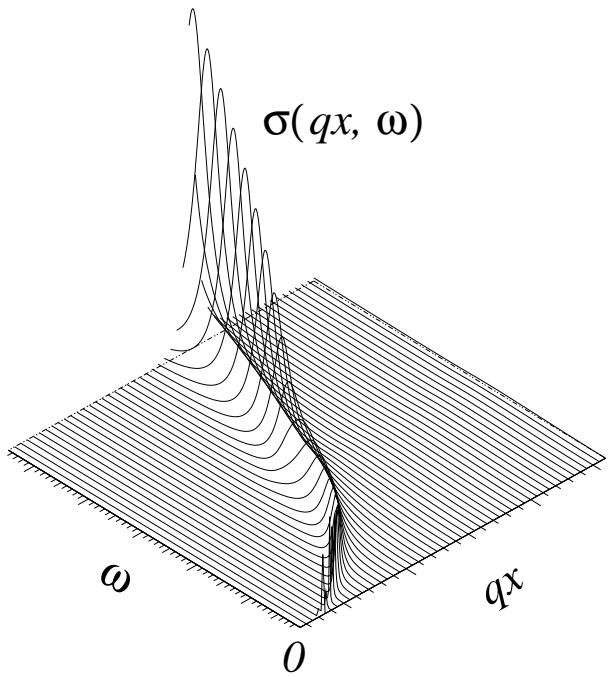}}
\caption{The conductivity  $\sigma_{xx}(q_x,\omega)$ (measured in arbitrary units) along the stripes in the disordered quantum smectic.}
\label{fig:sigma}
\end{figure}

In the clean system,  $\sigma_{xx}= Qq_x^4 \delta(\omega -
\sqrt{\kappa_\parallel Q} \vert q_x \vert^{5/2})$.  We see that
 there is no Drude peak for zero momentum but a  $\delta-$ function peak in
the conductivity at a  finite frequency $\omega_0(q_x) =\sqrt{\kappa_\parallel Q} \vert q_x \vert^{5/2})$.
In the presence of disorder,   akin to the clean system, the zero temperature
dc conductivity parallel to the stripes of the smectic vanishes for
all $q_x$.  From (\ref{conductivity}), we see
that  
the dynamical conductivity is  characterized by a pinning peak 
at non-zero frequency. 
Neglecting the dissipative term $I(\omega)$, we find a 
momentum dependent pinning frequency given by
\begin{equation}
\omega_{p}=  q_{x}^{2}\sqrt[]{Q\Sigma +Q\kappa_{\parallel } \vert q_{x}\vert }\simeq
 \sqrt{Q\Sigma q_{x}^{4}+\omega_0^2}
\label{pinning_freq}
\end{equation}
For $\kappa_{\parallel }|q_{x}|\ll \Sigma$, $\omega_{p}= \sqrt[]{Q\Sigma}q_{x}^{2}$.  In other disordered systems, the pinning frequency is basically fixed by $\Sigma$ and any small momentum dependence is just a shift in $\omega_p$.
However, in the case of the smectic, due to the $q_x$ dependent pre-factor,
$\omega_p$ can be made as small as possible by changing
$q_x$. 
The effect of the dissipative
term $I(i\omega_n)$ is to broaden the delta function at $\omega_p$
into a peak in conductivity. This peak is asymmetric  as in
various other quantum disordered systems\cite{chitra_wigner}. 
Finally,
for small $q_x$ and $\omega$, such that $ Q \Sigma q_x^4 \gg \omega^2$,
we obtain
$\sigma_{xx}(q_x,\omega) \simeq  \omega^\frac{15}{8}/\Sigma^2 $. 
Compared to other disordered systems where the low frequency
$q=0$ conductivity is quadratic in $\omega$, here  due to the
complicated scaling relations between $q_x$ and $\omega$, we find an exponent
which is slightly sub-quadratic. 
A plot of the conductivity as a function of $q_x$ and $\omega$ is on
Fig.~\ref{fig:sigma}. 
These $q$ dependent conductivities can be measured using the 
meander line techniques previously used to measure the 
microwave conductivity in
Wigner crystals.  To summarize, we see that disorder indeed has a big effect 
on ac conductivity.

Another quantity of interest is the threshold electric field for
depinning of the smectic.
Extending the Larkin-Ovchinnikov~\cite{larkin_ovchinnikov_pinning} arguments to
the present case, it is reasonable to assume that since the system is
pinned in the $x$ direction,  the threshold field
for depinning along this  direction is given by
$E_{Tx}= \kappa_\parallel R_a^{-1} l $  where $l$ is the average
width of a stripe. If the $y$ direction was free to slide, the pinning would
imply a strong non-linearity in the I-V characteristics for an applied field
in any arbitrary direction. 
Nonetheless, the result that the width fluctuations remain divergent along the
$y-$ direction does not  necessarily imply that there is no
threshold field for depinning in this direction. The behaviour of the
conductivity and threshold  in this
direction  will be dictated by   processes which facilitate  the
hopping  and tunneling of electrons between the stripes. Even in the
absence of disorder, 
the conductivity along the $y-$ direction is  through inter-stripe
tunneling events. These are not
included in our approach and depending on these processes,
$\sigma_{yy}(q=0,\omega=0)$ can be zero or finite. These   issues are
beyond the scope of the paper and  
will be
addressed 
in \cite{chitra_unpublished}.

To conclude, we have analyzed the pinning of quantum Hall stripes by
impurities using the Gaussian variational method. We find that at
$T=0$, the system is pinned and  insulating along the stripe direction
and
density  fluctuations exhibit  quasi-long range order akin to Bragg
glasses. Moreover, the dynamical conductivity in the parallel
direction
is strongly wave-vector dependent. It would also be of interest to perform a complementary
Functional Renormalization Group study of the QH stripes. To calculate $\sigma_{xy}$ and $\sigma_{yy}$
 and make a full comparison with experiments, one necessarily has to include
 processes which enable electrons to coherently  hop between stripes. This, 
however, is beyond the realm of the Gaussian Variational Method.    

\acknowledgments
We thank  T. Giamarchi  for discussions and for a careful reading
of the manuscript.


\end{document}